\documentclass[11pt]{article}
\usepackage{graphicx}
\usepackage{epstopdf}

\begin{document}

\def\xslash#1{{\rlap{$#1$}/}}
\def \p {\partial}
\def \dd {\psi_{u\bar dg}}
\def \ddp {\psi_{u\bar dgg}}
\def \pq {\psi_{u\bar d\bar uu}}
\def \jpsi {J/\psi}
\def \psip {\psi^\prime}
\def \to {\rightarrow}
\def\bfsig{\mbox{\boldmath$\sigma$}}
\def\DT{\mbox{\boldmath$\Delta_T $}}
\def\xit{\mbox{\boldmath$\xi_\perp $}}
\def \jpsi {J/\psi}
\def\bfej{\mbox{\boldmath$\varepsilon$}}
\def \t {\tilde}
\def\epn {\varepsilon}
\def \up {\uparrow}
\def \dn {\downarrow}
\def \da {\dagger}
\def \pn3 {\phi_{u\bar d g}}

\def \p4n {\phi_{u\bar d gg}}

\def \bx {\bar x}
\def \by {\bar y}

\begin{center}
{\Large\bf Gluon GPDs and Exclusive Photoproduction of a Quarkonium in Forward Region  }
\par\vskip20pt
Z.L. Cui$^{1,2}$,  M.C. Hu$^{1,2}$ and J.P. Ma$^{1,2,3}$    \\
{\small {\it
$^1$ Institute of Theoretical Physics, Chinese Academy of Sciences,
P.O. Box 2735,
Beijing 100190, China\\
$^2$ School of Physical Sciences, University of Chinese Academy of Sciences, Beijing 100049, China\\
$^3$ School of Physics and Center for High-Energy Physics, Peking University, Beijing 100871, China
}} \\
\end{center}
\vskip 1cm
\begin{abstract}
Forward photoproduction of $J/\psi$  can be used to extract Generalized Parton Distributions(GPD's) of gluons. 
We analyze the process at twist-3 level and  study relevant classifications of twist-3 gluon GPD's.  At leading power or twist-2 level the produced $J/\psi$ is transversely polarized.  We find that at twist-3 the produced $J/\psi$ is longitudinally polarized. 
Our study shows that in high energy limit the twist-3 amplitude is only suppressed by the inverse power of the heavy quark mass relatively to the twist-2 amplitude.   This indicates that the power correction to the cross-section 
of unpolarized $J/\psi$ can have a sizeable effect.  
We have also derived the amplitude of the production of $h_c$ at twist-3, but the result contains end-point singularities.  
The production of other quarkonia has been briefly discussed.

\vskip 5mm
\noindent
\end{abstract}
\vskip 1cm

\par

Properties of hadrons are described by their matrix elements of QCD operators.
These matrix elements are nonperturbative. They can be calculated only with nonperturbative methods, or extracted from experimental data.  A class of such matrix elements are    
Generalized Parton Distributions(GPD's) introduced in \cite{DMGPD, AVRGPD}.   In \cite{DVCSJi} it has been shown 
that one can obtain the quark- or gluon contributions to the proton spin from quark- or gluon 
twist-2 GPD's, respectively.  Therefore, studies of GPD's and relevant processes both in theory and experiment will 
help to solve the so-called proton spin crisis and provide more information about hadron's inner structure. 

\par  
There is a class of exclusive processes which can be used to extract GPD's. They can be extracted from Deeply Virtual Compton Scattering(DVCS)
in the forward limit, as suggested in \cite{DVCSJi}. 
Besides of DVCS,  processes 
with a light hadron or a quarkonium instead of the real photon in the final state of DVCS can be described with 
GPD's.  With the leading power approximation the amplitudes of the processes are given by convolutions of twist-2 GPD's 
with perturbative coefficient functions. The properties 
of twist-2 GPD's have been studied extensively(See reviews in \cite{DMGPDR,MDI,BeRa}).   From twist-2 GPD's one can only obtain 
the sum of the spin part and the orbital-angular momentum part of quarks or gluons. 
Recently, it has been shown in \cite{JXY,HaYo} that 
one can obtain each orbital angular momentum part from twist-3 GPD's individually.  Therefore, it is important to study twist-3 GPD's and how to extract them from experiment. 

\par     
In general twist-3 GPD's only appear in power-suppressed contributions. In this work we study the power-suppressed contributions of photoproduction 
of $J/\psi$ or $\Upsilon$.  The amplitude at the leading power of $J/\psi$- or $\Upsilon$ 
photoproduction   
has been given in \cite{KKMZ}.  At the leading power, only  a part of gluonic twist-2 GPD's is involved.  The produced 
quakonium at twist-2 level is only transversely polarized.  At the next-to-leading power, the quarkonium is longitudinally polarized. 
If one can measure the polarization,  then twist-3 GPD's can be accessed directly. It is interesting to note that 
the production rate of longitudinally polarized $J/\psi$- or $\Upsilon$ can be large because the polarization vector is proportional to the energy in contrast to the transverse polarization vectors which are constant. 
This can enhance the power-suppressed contributions in the high energy limit. This motivates us to study the process at twist-3 level in this work.   
 
\par 
We consider the process 
\begin{equation} 
   \gamma (q) + h(p) \to J_Q (k) +  h(p'),
\label{Eproc}
\end{equation}     
where the quarkonium is denoted as $J_Q$. We will consider the case where $J_Q$ is a $^3 S_1$ state of a $Q\bar Q$ state with $Q=c,b$ or a $^1P_1$ state.  The $^3S_1$ state corresponds to $J/\psi$ or $\Upsilon$, while the $^1P_1$ state corresponds to $h_{c,b}$.   In Eq.(\ref{Eproc}) the initial- or final hadron $h$ is a proton or a spin-1/2 hadron. 
The kinematical variables of the process are given by $s=(p+q)^2$ and $t= (p'-p)^2$.  This process with $J/\psi$
at leading twist 
has been studied extensively in, e.g., \cite{ISSK, VMJ,KKMZ} and references therein.    
\par 
We will use the  light-cone coordinate system, in which a
vector $a^\mu$ is expressed as $a^\mu = (a^+, a^-, \vec a_\perp) =
((a^0+a^3)/\sqrt{2}, (a^0-a^3)/\sqrt{2}, a^1, a^2)$ and $a_\perp^2
=(a^1)^2+(a^2)^2$. We define two light cone vectors $n^\mu = (0,1,0,0)$ and  $l^\mu =(1,0,0,0)$. The transverse metric is given by $g_\perp^{\mu\nu} = g^{\mu\nu} - n^\mu l^\nu - n^\nu l^\mu$. We will also need the transverse  antisymmetric tensor which is given by   
$\epsilon_\perp^{\mu\nu} =  \epsilon^{\alpha\beta \mu\nu } l_\alpha n_\beta$ with $\epsilon_\perp^{12}=1$.  We take a frame 
in which the momenta in Eq.(\ref{Eproc}) are given by: 
\begin{eqnarray} 
 p^\mu = (p^+, p^-,  -\frac{1}{2} \vec \Delta_\perp), 
\quad p'^\mu = (p'^+, p'^-, \frac{1}{2} \vec \Delta_\perp), \quad 
   q^\mu = (q^+,q^-,\frac{1}{2} \vec \Delta_\perp),  
\label{FRAME}    
\end{eqnarray}
with $\Delta^\mu = (p' -p)^\mu$. 
In the forward region the components of $\Delta^\mu$ except $\Delta^+$ are small, they scale like:
\begin{equation} 
\Delta_\perp^\mu \sim Q (\lambda,\lambda), \quad \Delta^- \sim p^-\sim p'^- \sim q^+ \sim Q\lambda^2
\end{equation}   
with $\lambda \ll  1$. $Q$ is a generic large scale of the process. It can be the heavy quark mass or $\sqrt{s}$.   In this kinematical region the momentum of $J_Q$ 
is:
\begin{equation} 
 \quad k_\perp^\mu =-\frac{1}{2}\Delta_\perp^\mu, \quad k^+ = -\Delta^+ + {\mathcal O}(Q \lambda^2), \quad k^- = q^-  + {\mathcal O}(Q\lambda^2). 
\end{equation}
The produced quarkonium moves almost in the direction of the initial photon. 

\par
For the process QCD factorization is expected in which the nonperturbative effect related to the quarkonium is included in matrix elements of nonrelativistic QCD(NRQCD)
and that of the initial hadron is included in various GPD's. 
The produced quarkonium mainly consists of a heavy quark $Q$ and an heavy antiquark $\bar Q$, the relative velocity $v$ between $Q$ and $\bar Q$ is small. One can make a small velocity expansion and use NRQCD factorization in \cite{nrqcd} for the quarkonium.  We take the leading order of $v$. At the leading order a quarkonium is 
a bound state of $Q\bar Q$. For our calculation one can use a wave function $\psi$ to project 
the relevant state of the produced $Q\bar Q$ pair. The projection is standard. For $J/\psi$ or $\Upsilon$ the derived amplitude will be proportional to the wave function at the origin. This can be replaced with the corresponding NRQCD matrix element defined in \cite{nrqcd}.      

\par
 
Before going to analyse the process, we introduce twist-3 parton GPD's and discuss their properties. 
We define the momentum $\bar P$ and the variable $\xi$ as:  
\begin{equation} 
    \bar P = \frac{p+p'}{2}, \quad  \xi =\frac{p^+ -p'^+}{ p^+ + p'^+} = \frac{-\Delta \cdot n}{2 \bar P\cdot n}.   
\end{equation}     
The twist-3 gluonic GPD's are given by the following matrix elements:
\begin{eqnarray}
M_D^{\mu\nu\rho} (x_1,x_2, \Delta ) &=&   \int \frac{d\lambda_1 d\lambda_2 }{(2\pi)^2} e^{i\bar P^+ \lambda_1 (x_1+x_2)/2
   + i\bar P^+ \lambda_2 (x_2-x_1)}  
 \nonumber\\   
   &&    
   \langle p',\lambda' \vert  G^{a, +\mu}(-\lambda_1n/2) 
     \tilde D_{\perp ab} ^\nu (\lambda_2 n ) \hat G^{b,+\rho} (\lambda_1 n/2 )    \vert p,\lambda \rangle,
\nonumber\\
M_F^{\mu\nu\rho} (x_1,x_2,\Delta ) &=&  i  f^{abc}  \frac{g_s}  {\bar P^+}  \int \frac{d\lambda_1 d\lambda_2 }{(2\pi)^2} e^{i\bar P^+ \lambda_1 (x_1+x_2)/2 
   + i\bar P^+ \lambda_2 (x_2-x_1)}
\nonumber\\   
   && \langle p' \vert  G^{a,+\mu } (-\lambda_1n/2 )     G^{b, +\nu}(\lambda_2 n )  G^{c,+\rho}(\lambda_1n/2)   \vert p \rangle, 
\nonumber\\
   M_\partial^{\mu\nu\rho} (x,\Delta) &=&   \frac{1 }{\bar P^+}  \int \frac{d\lambda}{2\pi} e^{ i\bar P^+ \lambda x }
 \langle p',\lambda' \vert    G^{a,+\mu} (-\lambda n/2 ) \tilde \partial^{ \nu}     G^{a,+\rho } (\lambda n/2)    \vert p,\lambda \rangle,      
\label{GGPD}    
\end{eqnarray}
where the derivative $\tilde \partial^\mu$ is defined as:
\begin{equation} 
    f^\dagger (x) \tilde \partial^\mu g(x) = \frac{1} {2}\biggr [  f^\dagger (x) \partial^\mu g(x)  - \biggr (\partial^\mu f(x) \biggr)^\dagger g(x)  \biggr ]  
\end{equation} 
for arbitrary functions $f(x)$ and $g(x)$. The covariant derivative $D^\mu$ is given by 
$D^{\mu} (x) = \partial^\mu + i g_s G^\mu (x)$. $\tilde D^\mu$ is defined with $D^\mu$ in the similar way 
of $\tilde \partial^\mu$.  All Lorentz indices $\mu,\nu$ and $\rho$ are transverse. 
There is another twist-3 matrix element of gluonic GPD's, which is obtained by replacing 
$f^{abc}$ in $M_F^{\mu\nu\rho} (x_1,x_2,\Delta )$ by $d^{abc}$. This matrix element is irrelevant for the 
process in Eq.(\ref{Eproc}). But it will be relevant for the process when the produced quarkonium is with the quantum number $C=+$.  

\par 
The definitions in Eq.(\ref{GGPD}) are given in the light-cone gauge $n\cdot G=0$. In other gauges one has to insert 
gauge links between field operators to make the definitions gauge invariant.  
These gauge links are built with:
\begin{equation} 
  {\mathcal L}(x) = P \exp\biggr \{ -i g_s \int_0^\infty d\lambda G^+(\lambda n +x) \biggr \}. 
\end{equation} 
It is noted that in gauges other than the light-cone one, the definition of $M_\partial^{\mu\nu\rho}$ 
is given by replacing $\tilde \partial^\nu$ with ${\mathcal L}^\dagger (-\lambda n/2) \tilde \partial^\nu {\mathcal L} (\lambda n/2)$ in Eq.(\ref{GGPD}).  Because the derivative also acts on gauge links, the symmetry of time-reversal does not 
give any constraint for $M_\partial^{\mu\nu\rho}$.  
The defined matrix elements are not independent.  Using the identity  
\begin{equation} 
{\mathcal L}(y n) D_\perp ^\mu (y n) {\mathcal L}^\dagger  (yn) F 
  = \partial_\perp^\mu F - i g_s \int^{\infty}_0 d\lambda \biggr ( {\mathcal L} G^{+\mu} {\mathcal L}^\dagger \biggr ) 
      ((\lambda+y) n) F, 
\label{IDT}
\end{equation}
for an arbitrary function $F$  one can derive the relation: 
\begin{eqnarray} 
M_D^{\mu\nu\rho} (x_1,x_2,\Delta) &=& \frac{1}{x_1-x_2 +i\varepsilon} M_F^{\mu\nu\rho} (x_1,x_2.\Delta) + \delta (x_2-x_1)  M_\partial^{\mu\nu\rho} (x_1,\Delta),   
\label{RMG}  
\end{eqnarray} 
where the factor $i\varepsilon$ in the denominator appears because the integration over $\lambda$ in the identity in Eq.(\ref{IDT}) 
is from $0$ to $\infty$. 
Therefore, only two of the three matrix elements in Eq.(\ref{GGPD}) are independent.    
\par 
In general, each matrix element 
with three transverse indices can be parameterized with scalar functions or GPD's. 
By taking Parity symmetry into account the number of such functions of each matrix element can not be larger than 16 for a spin-1/2 hadron. 
However, 
in this work, we will only encounter those matrix elements with two indices contracted with $g_\perp^{\mu\nu}$. 
There are three possible contractions.   
We consider first the contraction with the first- and third Lorentz index. We define the contracted matrix elements and 
give their parameterization as:   
\begin{eqnarray}
   G^\mu_{F,D} (x_1,x_2,\Delta) &=& g_{\perp\nu\rho} M_{F,D} ^{\rho\mu\nu} (x_1,x_2,\Delta), \quad, G^\mu_{\partial} (x,\Delta) = g_{\perp\nu\rho} M_{\partial} ^{\rho\mu\nu} (x,\Delta), 
 \nonumber\\  
   G^\mu_{\partial} (x,\Delta)  &=&  \bar u (p') \biggr [ i  \gamma^+ \frac{\Delta_\perp^\mu}{\bar P^+} H_1  + \frac{\sigma^{+\mu} m}{\bar P^+}  H_2  
 + \frac{\Delta_{\perp\rho}}{\bar P^+} \epsilon^{+-\mu\rho} \gamma^+ \gamma_5  H_3  
  + \frac{\Delta_\perp^\mu \sigma^{+\rho}\Delta_{\perp\rho}}{m \bar P^+} H_4
  \biggr ] u(p), 
\nonumber\\   
   G^\mu_{F} (x_1,x_2, \Delta )  &=&  \bar u (p') \biggr [ i  \gamma^+ \frac{\Delta_\perp^\mu}{\bar P^+} G_1  + \frac{\sigma^{+\mu} m}{\bar P^+}  G_2  
 + \frac{\Delta_{\perp\rho}}{\bar P^+} \epsilon^{+-\mu\rho} \gamma^+ \gamma_5  G_3  
  + \frac{\Delta_\perp^\mu \sigma^{+\rho}\Delta_{\perp\rho}}{m \bar P^+} G_4
  \biggr ] u(p), 
\nonumber\\
  G^\mu_{D} (x_1,x_2, \Delta ) &=& \bar u (p') \biggr [ i  \gamma^+ \frac{\Delta_\perp^\mu}{\bar P^+} F_1  + \frac{\sigma^{+\mu} m}{\bar P^+}  F_2  
 + \frac{\Delta_{\perp\rho}}{\bar P^+} \epsilon^{+-\mu\rho} \gamma^+ \gamma_5  F_3  
  + \frac{\Delta_\perp^\mu \sigma^{+\rho}\Delta_{\perp\rho}}{m \bar P^+} F_4
  \biggr ]  u(p)    
\label{PAGPDG1}          
\end{eqnarray}
where the scalar functions $G_i$, $F_i$ and $H_i$ depend on momentum fractions, $\xi$ and $t$. 
There is a freedom by choosing different parameterizations. The parameteriztion in the above  have the advantage that all spinor products $\bar u(p') \Gamma u(p)$ with 
$\Gamma = \gamma^+$, $\sigma^{+\mu}$ and $\gamma^+\gamma_5$ are at the order of ${\mathcal O}(\lambda^0)$.    
There are constraints for these functions or GPD's from hermiticity, Bose-symmetry and symmetry of time-reversal. 
All $G_i$ and $F_i$ are real and have the properties: 
\begin{eqnarray} 
  && G_3 (x_1,x_2,\xi,t) = - G_3 (x_2,x_1,-\xi,t), \quad G_{1,2,4}(x_1,x_2,\xi,t) = G_{1,2,4}(x_2,x_1,-\xi,t), 
\nonumber\\
  && F_3 (x_1,x_2,\xi,t)= F_3 (x_2,x_1,-\xi,t), \quad  F_{1,2,4} (x_1,x_2,\xi,t) = -F_{1,2,4}(x_2,x_1,-\xi,t), 
\nonumber\\
  && G_{i} (x_1,x_2,\xi,t) = -G_{i} (-x_2,-x_1, \xi,t),\quad F_i (x_1,x_2,\xi,t)= -F_i (-x_2, -x_1,\xi,t), \quad i=1,2,3,4. 
\label{PGF}     
\end{eqnarray}
It is noted that the symmetry of time-reversal does not give any constraint for $G_\partial^{\mu}$. The functions $H_i$ 
are complex in general.  The properties for $H_i$ are:
\begin{eqnarray} 
  H_{1,2,4}^* (x,\xi,t) = -H_{1,2,4}(x,-\xi,t),\quad H_3^*(x,\xi,t)=H_3 (x,-\xi,t), \quad H_{1,2,3,4} (x,\xi,t) = - H_{1,2,3,4} (-x,\xi,t). 
\end{eqnarray}
By using the identity  
\begin{equation}
 \frac{1}{x_1 -x_2 +i\varepsilon} = P\frac{1}{x_1-x_2} - i\pi \delta (x_1-x_2) 
\end{equation}
with $P$ standing for the principle-value prinscription and by noting that $F_{1,2,3,4}$ and $G_{1,2,3,4}$ 
are real and $H_{1,2,3,4}$ are complex, we obtain the relations among these GPD's   
from the relation in Eq.(\ref{RMG}):
\begin{eqnarray} 
F_{i} (x_1,x_2,\xi,t) &=& P\frac{1}{x_1-x_2} G_i (x_1,x_2,\xi,t) + \delta (x_2-x_1) {\rm Re} H_i (x_1,\xi,t),  
\nonumber\\
   G_i (x,x, \xi, t) &= & \frac{1}{\pi} {\rm Im} H_i (x,\xi,t).   
\label{LGFH}    
\end{eqnarray} 
In the limit of  $\Delta^\mu\to 0$, the GPD $G_2$, $F_2$ and $H_2$  become gluonic ETQS matrix elements 
relevant for single transverse-spin asymmetries in inclusive processes\cite{EFTE, QiuSt}. An interesting observation
of the limit  
is made in \cite{JXY,HaYo} that  one of moments of $F_3$ is the orbital-angular momentum contribution of gluons 
and that of $H_3$ can be interpreted as the canonical orbital-angular momentum contribution in the light-cone gauge 
as discussed in \cite{JM}.  It is noticed from Eq.(\ref{PGF},\ref{LGFH}) that $H_3$ becomes real in the limit by assuming 
$G_3$ is a continuous function of $\xi$.

\par
The matrix elements contracted in Eq.(\ref{GGPD}) with $g_{\perp\mu\nu}$ or $g_{\perp\nu\rho}$ are not independent. 
We have from Bose-symmetry:
\begin{equation} 
   g_{\perp\nu\rho} M^{\mu\nu\rho}_{D} (x_1,x_2, \Delta ) = -g_{\perp\nu\rho} M^{\nu\rho\mu}_{D} (-x_2, -x_1, \Delta ), 
   \quad g_{\perp\nu\rho} M^{\mu\nu\rho}_{\partial } (x, \Delta ) = -g_{\perp\nu\rho} M^{\nu\rho\mu}_{\partial} (-x, \Delta ).
\label{CT2}     
\end{equation}  
Since $M_F^{\mu\nu\rho}$ is defined with three field-strength-tensor operators,  
all three contractions give essentially the same GPD's
\begin{equation} 
   g_{\perp\nu\rho} M^{\mu\nu\rho}_{F} (x_1,x_2, \Delta ) = -g_{\perp\nu\rho} M^{\nu\rho\mu}_{F} (-x_2, -x_1, \Delta )  = - g_{\perp\nu\rho} M^{\nu\mu \rho}_{F} (x_1,x_1-x_2+ \xi,\Delta).  
\end{equation} 
We denote the matrix elements and parameterize them as:  
\begin{eqnarray} 
   \tilde G_D^{\mu} (x_1,x_2,\Delta) &=&  g_{\perp\rho\nu} M_D^{\mu\rho\nu}(x_1,x_2,\Delta), \quad \tilde G_\partial^\mu (x,\Delta)  = g_{\rho\nu} M_\partial^{\nu\rho\mu} (x,\Delta), 
\nonumber\\   
         \tilde G_\partial^\mu (x,\Delta)  &=& \bar u (p') \biggr [ i  \gamma^+ \frac{\Delta_\perp^\mu}{\bar P^+} \tilde H_1  + \frac{\sigma^{+\mu} m}{\bar P^+}  \tilde H_2  
 + \frac{\Delta_{\perp\rho}}{\bar P^+} \epsilon^{+-\mu\rho} \gamma^+ \gamma_5  \tilde H_3  
  + \frac{\Delta_\perp^\mu \sigma^{+\rho}\Delta_{\perp\rho}}{m \bar P^+} \tilde H_4
  \biggr ] u(p), 
\nonumber\\  
    \tilde G_D^{\mu} (x_1,x_2,\Delta)  &=& \bar u (p') \biggr [ i  \gamma^+ \frac{\Delta_\perp^\mu}{\bar P^+}  \tilde F_1  + \frac{\sigma^{+\mu} m}{\bar P^+}  \tilde F_2  
 + \frac{\Delta_{\perp\rho}}{\bar P^+} \epsilon^{+-\mu\rho} \gamma^+ \gamma_5  \tilde F_3  
  + \frac{\Delta_\perp^\mu \sigma^{+\rho}\Delta_{\perp\rho}}{m \bar P^+} \tilde F_4
  \biggr ]  u(p).    
\end{eqnarray}
From symmetries, the functions $\tilde F_i$ are real  and the functions $\tilde H_i$ are complex. 
They have the properties:
\begin{eqnarray} 
&& \tilde F_3 (x_1,x_2, \xi,t) = -  \tilde F_3 (-x_1,- x_2, -\xi,t), \quad  \tilde F_{1,2,4} (x_1,x_2, \xi,t) =  \tilde F_{1,2,4} (-x_1,- x_2, -\xi,t),  
\nonumber\\
&& \tilde H_3^* (x, \xi,t) = -  \tilde H_3 (-x, -\xi,t), \quad  \tilde H_{1,2,4}^* (x, \xi,t) =  \tilde H_{1,2,4}(-x, -\xi,t).  \end{eqnarray}
Again, these functions are not independent. They satisfy the relations: 
 \begin{eqnarray} 
\tilde F_{i} (x_1,x_2,\xi,t) &=& - P\frac{1}{x_1-x_2} G_i (x_1,x_1-x_2+\xi,\xi,t) + \delta (x_2-x_1) {\rm Re} \tilde H_i (x,\xi,t),  
\nonumber\\
      G_i (x_1,x_1-x_2+\xi,\xi,t) &=&  - \frac{1}{\pi} {\rm Im} \tilde H_i (x,\xi,t), 
\end{eqnarray} 
for $i=1,2,3,4$. These relations can be derived from Eq.(\ref{RMG}) similarly 
to those in Eq.(\ref{LGFH}).  
\par
Twist-3 matrix elements with quarks also appear in the twist-3 amplitudes of the considered processes at tree-level. 
There are three matrix elements: 
\begin{eqnarray} 
N_F^{\nu} (x_1,x_2,\Delta ) &=&  \int \frac{d\lambda_1}{2\pi} \frac{ d\lambda_2}{2\pi} e^{i\bar P^+ \lambda_1 (x_1+x_2)/2 
   + i\bar P^+ \lambda_2 (x_2-x_1)} 
 \langle p' \vert \bar \psi (-\lambda_1 n/2 ) \gamma^+ g_s G^{+\nu}(\lambda_2 n )  \psi (\lambda_1n/2)  \vert p \rangle,   
\nonumber\\
N_D^{\nu} (x_1,x_2,\Delta) &=&   \bar P^+  \int \frac{d\lambda_1}{2\pi} \frac{ d\lambda_2}{2\pi} e^{i\bar P^+ \lambda_1 (x_1+x_2)/2 
   + i\bar P^+ \lambda_2 (x_2-x_1)}
 \langle p' \vert \bar \psi  (-\lambda_1 n /2) \gamma^+  \tilde D_\perp^\nu (\lambda_2 n ) \psi (\lambda_1n/2)  \vert p \rangle,
\nonumber\\
   N_\partial^{\nu} (x, \Delta) &=& \int \frac{d\lambda}{2\pi}  e^{ i\bar P^+ \lambda x}
  \langle p' \vert \bar\psi (-\lambda  n /2)  \gamma^+ \tilde \partial_\perp^\nu  \psi (\lambda n/2)   \vert p \rangle. 
\end{eqnarray} 
Similarly, one can derive the relation between these three matrix elements by using the identity in Eq.(\ref{IDT}):
\begin{eqnarray} 
N_D^{\nu} (x_1,x_2,\Delta ) &=& \frac{1}{x_1-x_2 +i\varepsilon} N_F^{\nu} (x_1,x_2,\Delta ) + \delta (x_2-x_1) N_\partial^{\nu} (x_1,\Delta ).  
\label{RNFD}   
\end{eqnarray} 
Similar relation has been also derived in \cite{HaYo}, where $i\varepsilon$ in the denominator is missing. 
In our work, only the sum of all quark flavors is relevant. We define the following quark-gluon GPD's: 
\begin{eqnarray} 
\sum_q N_F^{\mu} (x_1,x_2) &=&\bar u (p') \biggr [ i  \gamma^+ \frac{\Delta_\perp^\mu}{\bar P^+}  N_1   + \frac{\sigma^{+\mu} m}{\bar P^+}  N_2  
 + \frac{\Delta_{\perp\rho}}{\bar P^+} \epsilon^{+-\mu\rho} \gamma^+ \gamma_5  N_3   
  + \frac{\Delta_\perp^\mu \sigma^{+\rho}\Delta_{\perp\rho}}{m \bar P^+} N_4 
  \biggr ]  u(p).   
\end{eqnarray}
The sum is over light flavors of quarks. In the limit of  $\Delta^\mu\to 0$,  one of moments of $N_3$ is the orbital-angular momentum contribution of quarks as observed in \cite{JXY, HaYo}. 
\par 
  For completeness of gluon GPD's we also include definitions of gluonic twist-2 GPD's. 
These GPD's have been discussed in detail  in \cite{MDI, BeRa}. They will appear in our amplitudes. We take the notation of \cite{MDI} for twist-2 gluon GPD's:
\begin{eqnarray} 
  {\mathcal F}_g &=& \frac{1}{\bar P^+} \int\frac{d\lambda}{2\pi} e^{ i x \bar P^+ \lambda} 
    \langle p' \vert g_{\perp\mu\nu} G^{+\mu} (-\frac{1}{2} \lambda n) G^{+\nu}  (\frac{1}{2} \lambda n ) \vert p \rangle  
\nonumber\\
     &=& -\frac{1}{2 \bar P^+} \bar u(p') \biggr [ \gamma^+ H_g (x, \xi, t) + \frac{i\sigma^{+\alpha}} {2 m} \Delta_\alpha 
     E_g (x,\xi,t) \biggr ] u(p), 
\nonumber\\
  \tilde {\mathcal F}_g    &=& \frac{i}{\bar P^+} \int\frac{d\lambda}{2\pi} e^{ i x \bar P^+ \lambda} 
    \langle p' \vert \epsilon_{\perp\mu\nu} G^{+\mu} (-\frac{1}{2} \lambda n) G^{+\nu} (\frac{1}{2} \lambda n ) \vert p \rangle
\nonumber\\
     &=& -\frac{1}{2 \bar P^+} \bar u(p') \biggr [ \gamma^+ \gamma_5 \tilde  H_g (x, \xi, t) + \frac{\gamma_5 \Delta^+ } {2 m} 
      \tilde E_g (x,\xi,t) \biggr ] u(p),
\nonumber\\
  {\mathcal F}^{\mu\nu }_{Tg} &=&  \frac{1}{2 \bar P^+} \int\frac{d\lambda}{2\pi} e^{ i x \bar P^+ \lambda} 
    \langle p' \vert  {\bf S} \biggr (   G^{+\mu } (-\frac{1}{2} \lambda n) G^{+\nu}  (\frac{1}{2}  \lambda n ) \biggr )   \vert p \rangle  
\nonumber\\
    &=& -\frac{1}{2\bar P^+} {\bf S} \biggr \{ \frac{\bar P^+ \Delta^\nu - \Delta^+ \bar P^nu}{2 m \bar P^+} \bar u(p') \biggr [ 
           i\sigma^{+\mu} H_{gT} (x,\xi,t) +\frac{ \bar P^+ \Delta^\mu -\Delta^+ \bar P^\mu}{m^2} \tilde H_{gT} (x,\xi,t) 
\nonumber\\
    && +\frac{\gamma^+ \Delta^\mu -\Delta^+ \gamma^\mu}{2 m} E_{gT} (x,\xi,t) + \frac{\gamma^+\bar P^\mu-\bar P^+ \gamma^\mu }{m} 
     \tilde E_{gT} (x,\xi,t) \biggr ] u(p)  \biggr \},  
\label{GT2}                             
\end{eqnarray}  
where $\mu$ and $\nu$ are transverse. The notation ${\bf S}(\cdots)$ implies that the tensors in $(\cdots)$ are symmetric and traceless.  There are in total eight twist-2 gluon GPD's. Their properties can be found in \cite{MDI, BeRa}.

\par 
 
\begin{figure}[hbt]
	\begin{center}
		\includegraphics[width=12cm]{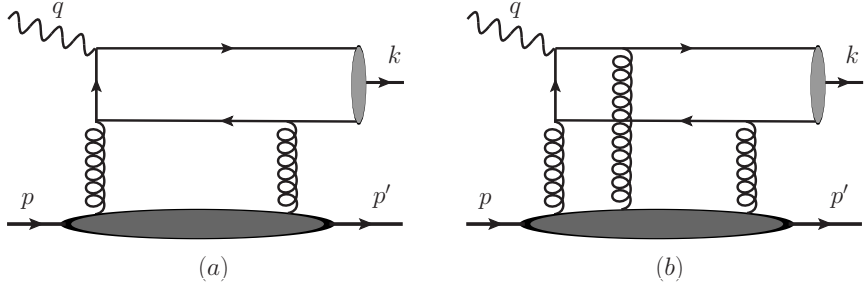}
	\end{center}
	\caption{  (a): One of the six diagrams for two-gluon exchanges. (b): One diagrams for three-gluon exchanges. Other diagrams are obtained through permutation.   }
	\label{P1}
\end{figure}

\par 
Now we turn to our derivation of  the twist-2- and twist-3 amplitude for the process in Eq.(\ref{Eproc}) at tree-level 
in the case that the quarkonium is a $^3 S_1$ $Q\bar Q$ state. 
We will perform the calculation mainly in the light-cone gauge $n\cdot G =G^+= 0$, but also partly in Feynman gauge 
for fixing some ambiguities. At tree-level, the quarkonium is produced through exchanges of gluons between 
the heavy quark pair and the initial hadron.  The two-gluon- or three-gluon exchanges are illustrated in Fig.1a or Fig.1b, 
respectively.  After projecting  out the $^3S_1$ state from the produced $Q\bar Q$ pair, the contribution from Fig.1a can be written as:
\begin{eqnarray} 
i {\mathcal T} \biggr\vert_{1a} = \frac{1}{2} \int d^4 k_1 d^4 k_2 {\mathcal M}^{\mu\nu} (k_1,k_2,\Delta) \int \frac{d^4 \xi_1}{(2\pi)^4} 
 \frac{d^4 \xi_2}{(2\pi)^4}  e^{i\xi_1 \cdot k_1+ i\xi_2 k_2} 
   \langle p' \vert G_\mu (\xi_1) G_\nu(\xi_2) \vert p \rangle,
\label{F1A}     
\end{eqnarray}       
where ${\mathcal M}^{\mu\nu} (k_1,k_2,\Delta)$ is essentially  the amplitude of $\gamma(q) + g(k_1) + g(k_2)  \to J_Q (k)$ after taking the trace of colors.  It is given by six diagrams.  To make power-expansion related to the light hadron, one notes that in the forward limit the gauge fields in the matrix element of Eq.(\ref{F1A}) and the momenta carried by these 
fields scale like:
\begin{equation} 
   G^\mu \sim Q (1,\lambda^2,\lambda,\lambda), \quad k_{1,2}^\mu \sim Q (1, \lambda^2,\lambda,\lambda). 
\label{PC}     
\end{equation}  
In the limit, the momentum transfer $\Delta^\mu$ also scales like $\Delta^\mu\sim Q (1, \lambda^2,\lambda,\lambda)$.  The twist-2- or leading power contribution of the amplitude is obtained by taking the leading order 
of the expansion in $\lambda$.  For this we need to expand ${\mathcal M}^{\mu\nu} (k_1,k_2,\Delta)$ and the matrix element in $\lambda$. In the light-cone gauge, the leading contribution of the matrix element is given when the two 
gluon fields carry transverse indices, i.e., $\mu,\nu=\perp$. ${\mathcal M}^{\mu\nu} (k_1,k_2,\Delta)$ at the leading order 
for $\mu,\nu=\perp$ is given by:
\begin{eqnarray} 
{\mathcal M}^{\mu\nu} (k_1,k_2,\Delta) &=&i  g_s^2 A_J 
 \biggr [  \frac{16 m_Q^2}{k^+ q^-} \frac{k_1^+ k_2^+}{ (k_2^+ - i\varepsilon) (k_1^+ - i\varepsilon)}
   g_\perp^{\mu\nu} \epsilon_\perp (q) \cdot \epsilon^*(k) \biggr ] 
  ( 1 +{\mathcal O}(\lambda^2) ),  
\nonumber\\
   && A_J = Q_Q e   \psi^* (0) \frac{(2m_Q)^{-3/2}}{\sqrt{N_c}},        
\end{eqnarray}      
where $\psi(0)$ is the non-relativistic wave function at the origin.  $\epsilon(q)$ and $\epsilon(k)$ are the polarization vector of the photon and $J_Q$, respectively. The factor $i\varepsilon$ in denominators 
comes from the quark propagators in Fig.1a. 
In the gauge $n\cdot G=0$, the twist-2 contribution comes only from two-gluon exchanges as in Fig.1a.  
It is straightforward to obtain the twist-2 amplitude denoted as ${\mathcal T}_T$:  
\begin{eqnarray}  
 {\mathcal T}_T  &=& 8 g_s^2 A_J 
      \epsilon_\perp (q) \cdot \epsilon^*(k)  
 \int d x  \frac{1}{( x-\xi + i\varepsilon)( x+\xi -i\varepsilon) }{\mathcal F}_g(x,\xi,t).  
\label{T2}      
\end{eqnarray}
This result is in agreement with that given in \cite{KKMZ}. We note here that at twist-2 the produced $J\psi$ or $\Upsilon$ is transversely polarized.  Using this amplitude 
one can extract from experimental data only the GPD $H_g$ and $E_g$ contained in ${\mathcal F}_g$.  

\par 
Now we consider the amplitude with longitudinally polarized $J/\psi$ or $\Upsilon$. This amplitude is power-suppressed. 
In contrast to the twist-2 amplitude ${\mathcal T}_T$,  there are several sources of contributions. 
There are contributions which involve twist-2 gluon GPD's and are explicitly power-suppressed. They are  called as dynamical twist-3 contributions. There are also 
contributions which involve twist-3 GPD's. One part of dynamical twist-3 contributions can be easily identified from Eq.(\ref{T2}). 
In the used frame given by Eq.(\ref{FRAME}), the polarization vector $\epsilon_L$ of the longitudinally polarized 
$J/\psi$  has a small transverse part at order of $\lambda$.  
To specify the polarization of $J/\psi$, one has to define the spin quantization axis properly. 
We take the so-called $s$-channel helicity frame by introducing the 4-vector $X^\mu = q+p$. The polarization 
vector $\epsilon_L^\mu$ of this frame is given by:
\begin{equation} 
  \epsilon_L^\mu (k)  = \frac{k\cdot X}{\sqrt{(k\cdot X)^2 - M^2 s}}
\biggr ( \frac{k^\mu}{M} -\frac{M}{k\cdot X}X^\mu \biggr).  
\end{equation}    
The vector $\epsilon_L$ given here is covariant with $k^2 =M^2$. The physical meaning of the vector $X^\mu$ is that in the rest frame of $J_Q$ the spin quantization direction is the direction of the 3-vector $-{\bf X}$.   
Keeping the leading order of $\lambda$, we have  
\begin{equation} 
\epsilon_L^\mu = (\epsilon_L^+,\epsilon^-_L,0,0) +\epsilon_\perp^\mu + {\mathcal O}(\lambda^2), 
 \quad \epsilon_L^- = \frac{k^-}{2 m_Q},\quad \epsilon_{L\perp}^\mu = \frac{\rho _L}{2 m_Q} k_\perp^\mu, \quad \rho_L = \frac{s+ 4m_Q^2}{s-4 m_Q^2}. 
\label{LPV} 
\end{equation}
The component $\epsilon_L^+$ is not needed to express our result. With the given $\epsilon_{L\perp}^\mu$ we obtain the part of dynamical twist-3 contributions from the transverse part of $\epsilon_L^\mu$ can be read off 
from Eq.(\ref{T2}):    
\begin{eqnarray}  
 {\mathcal T} \biggr\vert_{a,30} &=&  -  g_s^2 A_J  
     \frac{2}{ m_Q } \epsilon_\perp (q) \cdot \Delta_\perp \rho_L    
 \int d x  \frac{1}{( x-\xi + i\varepsilon)( x+\xi -i\varepsilon) }{\mathcal F}_g(x,\xi,t), 
\label{KL}     
\end{eqnarray}
which is power-suppressed indicated by the factor $\Delta_\perp$. 
\par 
In the gauge $n\cdot G=0$ the remaining twist-3 contributions come from two-gluon- and three-gluon exchanges as diagrams in  Fig.1.  We first study the contributions from Fig.1a.  It is nontrivial 
to find the twist-3 contributions from Fig.1a because Fig.1a also has the leading- or twist-2 contributions. 
With the power counting given in Eq.(\ref{PC}), one part of twist-3 contributions is given when one of the two gluon fields 
carries the $-$-index. In this case one should take the leading order of the $\lambda$-expansion for ${\mathcal M}^{\mu+} $. 
Another part comes from the next-to-leading order of the $\lambda$-expansion 
for ${\mathcal M}^{\mu\nu}$ with $\mu=\nu=\perp$. We denote this contribution of  ${\mathcal M}^{\mu+}$ 
as ${\mathcal M}_1^{\mu\nu}$.  After the expansion of ${\mathcal M}^{\mu\nu}$
we have: 
\begin{eqnarray}
{\mathcal M}^{\mu+} (k_1,k_2,\Delta)  &=&{\mathcal M}^{+\mu}(k_1,k_2,\Delta) =  i  g_s^2 A_J 
 \epsilon^\mu\biggr [  \frac{- 16 m^2_Q \epsilon_L^-   k_1^+ k_2^+ }{  q^- k^- (k_2^+ - i\varepsilon) (k_1^+ - i\varepsilon)} \biggr ] ( 1+{\mathcal O}(\lambda) ) ,    
\nonumber\\ 
{\mathcal M}^{\mu\nu}_1(k_1,k_2,\Delta)  &=& i  g_s^2 A_J  
 \biggr [  \frac{16 m^2_Q \epsilon_L^- \epsilon_\rho }{ k^+ q^- k^- (k_2^+ - i\varepsilon) (k_1^+ - i\varepsilon)}
   \biggr ( g_\perp^{\mu\nu} k_{1\perp}^\rho (k_2^+)^2  + g_\perp^{\mu\nu} k_{2\perp}^\rho (k_1^+)^2
\nonumber\\
  &&  - k_{1\perp}^\nu g_\perp^{\mu\rho} k_2^+ (k_1^+ + k_2^+) - k_{2\perp}^\mu g_\perp^{\nu\rho}  k_1^+ (k_1^+ + k_2^+) \biggr ) \biggr ],  
\label{CMP1}   
\end{eqnarray} 
with $\mu,\nu=\perp$. 
From ${\mathcal M}^{\mu+}$ we have the twist-3 contribution:
\begin{eqnarray}  
i {\mathcal T} \biggr\vert_{a,31}      &=& i g_s^2 A_J   
 \int  dk_1^+ d k_2^+  \biggr [  \frac{16 m^2_Q \epsilon_L^- \epsilon_\mu  }{q^- k^- (k_2^+ -  i\varepsilon) (k_1^+ - i\varepsilon)} \biggr ] \int \frac{ d\xi_1^- d\xi_2^-}{(2\pi)^2}e^{i\xi_1^-  k_1^+  + i\xi_2^- k_2^+}
\nonumber\\ 
   && \langle p' \vert G^{+\mu} (\xi_1^-n) G^{+-} (\xi_2^-n )  \vert p \rangle. 
\label{1A31}    
\end{eqnarray} 
It is noted that the field $G^-$ is not dynamically independent.  From QCD equation of motion in the gauge $G^+=0$ we
have the equation for $G^{+-}$:  
\begin{equation} 
   \partial^+ G^{a,-+}(x) =  - \biggr ( D_{\perp\mu} (x)  G^{\mu +} (x) \biggr )^a   + g_s J^{a,+}(x),  \quad J^{a,\mu} (x) = \sum_q \bar q (x) T^a \gamma^\mu q(x),           
\end{equation} 
and the solution is: 
\begin{eqnarray} 
   G^{a,+-} (x)     &=& i \int \frac{d\lambda d k^+}{2\pi} \frac{e^{-i\lambda k^+} }{k^+ +i\varepsilon} \biggr [ \biggr ( D_{\perp\mu} G^{+\mu}\biggr )^a (x +\lambda n) 
      + g_s J^{a,+}(x+\lambda n) \biggr ].   
\end{eqnarray} 
Using this result with little algebra we derive: 
\begin{eqnarray} 
i {\mathcal T}\biggr\vert_{a,31} &=& g_s^2 A_J\frac{16 m^2_Q \epsilon_L^- }{( k^-)^2} \epsilon_\rho  
 \int dk_1^+ d k_2^+  \biggr [ \frac{ 1 }{ (k_2^+ - i\varepsilon)^2 (k_1^+ - i\varepsilon)} \biggr ]
     \int \frac{ d\xi_1^- d\xi_2^-}{(2\pi)^2}e^{i\xi_1^-  k_1^+ + i\xi_2^- k_2^+} 
\nonumber\\
   &&   
   \biggr \{  \langle p' \vert     g_sG^{+\rho} (\xi_1^-n)  J^{+}(\xi_2^- n )  + \frac{i}{2} \Delta_{ \perp\nu}   G^{+\rho} (\xi_1^-n)  G^{+\nu} (\xi_2^- n) 
      \vert p \rangle 
\nonumber\\      
     && + g_{\mu\nu} \int d k_3^+ \int \frac{ d\xi_3^-}{2\pi} e^{ i k_3^+(\xi_3^--\xi_2^-) } \langle p' \vert G^{+\rho} (\xi_1^-n)  \tilde  D_{\perp}^\mu  (\xi_3^- n)  G^{+\nu}(\xi_2^- n) 
      \vert p \rangle \biggr \}.   
\end{eqnarray}
It is noted that the matrix element with $G^{+-}$ in Eq.(\ref{1A31}) contains not only twist-3 matrix element 
but also twist-2 matrix element. 

\par  
To the contribution from ${\mathcal M}^{\mu\nu}_1$ in Eq.(\ref{CMP1}),  we have to add the factor 
\begin{equation} 
\frac{k_1^+}{k_1^+-i\varepsilon} , \quad \frac{k_2^+}{k_2^+ - i\varepsilon} , 
\label{FAS} 
\end{equation} 
in order to express the result with matrix elements of field strength tensor operators in the light-cone 
gauge $n\cdot G=0$.  These factors can be different from $1$ under integrations over $k_{1,2}^+$. We will 
discuss these added factors later. The contribution reads: 
\begin{eqnarray} 
i {\mathcal T} \biggr\vert_{a,32}   &=&   g_s^2 A_J\frac{16 m^2_Q \epsilon_L^- }{k^+( k^-)^2}  
 \int dk_1^+ d k_2^+  \biggr [ \frac{ 1 }{ (k_2^+ -  i\varepsilon)^2 (k_1^+ - i\varepsilon)} \biggr ]
   \epsilon_\rho \int \frac{ d\xi_1^- d\xi_2^-}{(2\pi)^2}e^{i\xi_1^-  k_1^+  +  i\xi_2^- k_2^+} g_{\perp \mu\nu}
\nonumber\\
   &&   
   \langle p' \vert   \biggr [   k_1^+  G^{+\mu}(\xi_1^- n)
   \tilde  \partial_{\perp}^\rho  G^{+\nu}(\xi_2^-n)  + \frac{i}{2} k_1^+ \Delta_\perp^\rho G^{+\mu}(\xi_1^- n)
    G^{+\nu}(\xi_2^-n)   
\nonumber\\    
   &&  - k^+   G^{+\mu}(\xi_1^- n) \tilde \partial_\perp^\nu G^{+\rho} (\xi_2^- n) -\frac{i}{2} k^+ \Delta_{\perp\nu}  G^{+\mu}(\xi_1^- n)  G^{+\rho} (\xi_2^- n)  
  \biggr ]   \vert p \rangle. 
\label{1A32}                                     
\end{eqnarray}
This part contains not only twist-3 matrix elements but also those at twist-2 matrix elements. 
\par 
The contribution from Fig.1b is relatively easy to obtain, because of that its leading contribution is at twist-3 in the light-cone gauge.  We obtain:
\begin{eqnarray} 
i {\mathcal T} \biggr\vert_{b,3} &=&  g_s^3 A_J   \frac{16 m^2_Q \epsilon_L^{*-} }{(k^-)^2}  \int  d k_1^+ d k_2^+ d k_3^+  \frac{ -i^3 } {(k_2^+ - i\varepsilon)(k_3^+- i\varepsilon)}
 \cdot  \frac{  f^{a_1a_2a_3} g_{\perp\mu_1\mu_2} \epsilon_{\mu_3}}{(k_2^+ + k_3^+ -i\varepsilon)(k_1^+ +k_3^+ -i\varepsilon) } 
\nonumber\\
  &&  \int \frac{d \xi_1^-}{(2\pi)} 
 \frac{d\xi_2^-}{(2\pi)} \frac{d\xi_3^-}{(2\pi)} e^{i\xi_1^-  k_1^+ + i\xi_2^- k_2^+ + i\xi_3^- k_3^+ } 
   \langle p' \vert G^{a_1,+\mu_1} (\xi_1^-n) G^{a_2,+\mu_2}(\xi_2^-n)  G^{a_3,+\mu_3} (\xi_3^-n) \vert p \rangle. 
\label{1B3}        
\end{eqnarray}  
Comparing the definitions in Eq.(\ref{GGPD}),  this contribution can be expressed with the twist-3 matrix element 
$M_F^{\mu\nu\rho}$. It is noted that this contribution can also be expressed with $M_D^{\mu\nu\rho}$ by adding 
the contribution of the first term in the second line of Eq.(\ref{1A32}).  Since the three twist-3 matrix elements in Eq.(\ref{GGPD}) are not independent, there is the freedom to choose an independent set 
of these matrix elements. We will express 
our final results with $M_{D,\partial}^{\mu\nu\rho}$.  
\par 
In deriving our results in the above, we have worked with the light-cone gauge $G^+ =0$. In this gauge the gauge field 
operator in matrix elements can be converted to the field-strength tensor operator combined with its momentum:
\begin{equation} 
  \int d^4 \xi_i  e^{i \xi_i \cdot k_i } k_{i}^+ G^{\mu} (\xi_i) \to  i  \int d^4 \xi_i e^{i \xi_i\cdot k_i }  G^{+ \mu} (\xi_i), 
  \quad i=1,2.  
\end{equation} 
In our calculations the needed factors $k_{1,2}^+$ are from ${\mathcal M}^{\mu\nu}$ and from the upper part of 
Fig.1b., except for the contribution given in Eq.(\ref{1A32}), where we have added the factors in Eq.(\ref{FAS}). 
This introduces an ambiguity for the sign before $i\varepsilon$.
It can be checked by doing the calculation in Feynman gauge.  
We have calculated the twist-3 contributions of Fig.1a in this gauge, and found that the gauge fields 
in these contributions always appear in the combined form for $\mu\neq +$:
\begin{equation} 
  \int d^4 \xi_i  e^{i \xi_i \cdot k_i }  \biggr [  G^{\mu} (\xi_i)  - G^+(\xi_i) \frac{ k_{\perp i}^\mu}{ k_i^+ -i\varepsilon} \biggr ], 
    \quad i=1,2.  
\end{equation}        
This combination gives the form of field-strength tensor operators at the leading order of $g_s$. This 
indicates that the sign of $i\varepsilon$ in Eq.(\ref{FAS}) is correct. We have also 
calculated the contribution of Fig.1b with one gluon field carrying the $+$-index in Feynman gauge. This contribution 
can be divided into two parts. One part gives the above combination, therefore is included in  field-strength tensor operators in Eq.(\ref{1B3}) for Fig.1b.  If we take the contribution from Fig.1a given in Eq.(\ref{1A32}) as gauge invariant, 
it implies that there is the contribution from three-gluon exchanges with one gluon field carrying the $+$-index
and other two carrying a transverse index.  This contribution comes partly from the non-linear term of field-strength tensor operators and partly from gauge links between operators.  The remaining part from Fig.1b with one gluon field carrying the $+$-index just gives this contribution. This gives an important check for gauge invariance of our results.   
\par 
Our result for the amplitude at twist-3, denoted as ${\mathcal T}_L$, is the sum of the amplitudes given in Eq.(\ref{KL}, \ref{1A31},\ref{1A32},{\ref{1B3}).  It is:  
\begin{eqnarray} 
i{\mathcal T}_L    &=&  2 g_s^2A_J\frac{\epsilon_\rho }{ m_Q }  \biggr \{   \int d x  \frac{1}{ (x+\xi-i\varepsilon)^2 (\xi-x-i\varepsilon)}        \biggr [  i  \Delta_\perp^\rho ( \xi-x   - \rho_L (x +\xi )  )  {\mathcal F} _g(x,\xi,t) 
\nonumber\\  
    &&  + 2 \xi \epsilon_\perp^{\mu\rho} \Delta_\mu  \tilde {\mathcal  F}_g(x,\xi,t)   \biggr ]   
      + 4\xi    
 \int d x_1 d x_2   \biggr [  \frac{ \delta (x_1-x_2) \tilde G_\partial^\rho  (- x_1,\Delta )  }{ ( x_1+\xi - i\varepsilon)^2 (\xi-x_1 - i\varepsilon)} 
 \nonumber\\      
     && +    \frac{  \tilde G_D^\rho  (x_1,x_2,\Delta ) }{  ( x_2+\xi - i\varepsilon)^2 (\xi-x_2 - i\varepsilon)}      
  +      \frac{G_D^\rho  (x_1,x_2,\Delta) }{  (\xi+x_1 - i\varepsilon) (\xi-x_1 - i\varepsilon)(\xi+x_2 -i\varepsilon ) }      
\nonumber\\
  && +   \frac{  \sum_q N_F^{\rho}(x_1,x_2,\Delta) }{ (2\xi - x_2 + x_1- i\varepsilon)^2 (x_2-x_1 - i\varepsilon)}  \biggr ) 
 \biggr ]   \biggr \}. 
\label{T3}  
\end{eqnarray} 
This result is complicated when we express it with various GPD's. Totally, 20 GPD's or scalar functions are involved 
with the parameterization of twist-2 and twist-3 matrix elements given before. This makes  the extraction of GPD's 
from experiment complicated. Unless one can use some nonperturbative methods to calculated these GPD's, 
it seem impossible to make reliable predictions for experiment.

\par 
Before giving a discussion about our results, we notice that our amplitudes in Eq.(\ref{T2}, \ref{T3}) 
are  $U_{em}(1)$-gauge invariant only at the leading order of $\lambda$.
Since in our results only the transverse part of the polarization vector 
of the initial photon is involved, the amplitudes are not exactly zero by replacing the polarization vector with the 
photon momentum $q^\mu$. After replacement the amplitudes are in fact suppressed 
by one power of $\lambda$ because $q_\perp^\mu$ is at order of $\lambda$. Hence, 
$U_{em}(1)$-gauge invariance of our result holds only at the leading order of $\lambda$.

\par 
One may expect that the amplitude can be simplified in the high energy  limit, i.e., $s \gg m_Q^2$.
 In the process $\xi$ is fixed as 
$\xi=2m_Q^2/(s-2 m_Q^2)  (1+ {\mathcal O}(\lambda)$.
Formally, one sets $\xi=0$ to significantly simplify the twist-3 amplitude in the limit $s \gg m_Q^2$. 
But, this will results in a divergent contribution. This can be seen from the contribution from 
the GPD $H_g(x,\xi,t)$ contained in ${\mathcal F}_g$. In the limit $\Delta^\mu \to 0$ $H_g$ is related to the gluon 
parton distribution $g(x)$ as $ H_g (x,0,0)= x g(x)$. For $x\to 0$ $g(x)$ behaves as $x^{-\alpha}$ with $\alpha >0$. 
The pole contribution from $H_g$ in the first line of Eq.(\ref{T3}) is expected for $\xi\to 0$ as $\xi^{-\alpha}$, as discussed in \cite{MaPi}.  
The behaviour of GPD's in the region $\vert x\vert \sim \xi$ 
is crucial for the limit. It is clear that the integration regions with $\vert x\vert \sim \xi$ and $\vert x_{1,2}\vert \sim \xi$ are the most important ones. 
If one assumes all GPD's have similar behaviours in these regions, one may only neglect the last two terms by power 
counting of $\xi$. In general, it is not clear if the amplitude will be simplified in the limit. But, the relative order of 
importance between twist-2- and twist-3 amplitude can be estimated.      

\par 
By simply comparing the contribution from $H_g$ in twist-2- or twist-3 contribution, one can already conclude that 
the twist-3 amplitude is suppressed  only by a factor $\Delta_\perp/m_Q$. For $J/\psi$ $m_Q$ is $m_c$ around $1$GeV. Therefore, the suppression is not strong.  The weak suppression  has an important impact on extracting twist-2 gluon GPD's by 
using the leading amplitude given in Eq. (\ref{T2}), in which $J/\psi$ is only transversely polarized.  
If the polarization of$J/\psi$ is not measured or summed in experiment, 
then there will be a substantial contribution of longitudinally polarized $J/\psi$ in the measured differential cross-sections. This can result in that the extracted twist-2 GPD's by using  the leading amplitude are not accurate and 
contain significant effects of higher-twists. This is an important observation of our work.  

\par     
From the production of transversely polarized $J/\psi$ or $\Upsilon$  one can only extract 
the gluon GPD's $E_g$ and $H_g$.  It is possible to obtain information of other  twist-2 
GPD's through production of quankonium with different quantum numbers. At leading power or twist-2 level, 
the exchanged two gluons in Fig.1a can only be in a state with the quantum number $C=+$.  Therefore, 
$h_c$ can be produced at twist-2.  Other quarkonia like $\eta_c$ and $\chi_{cJ}$ with $J=0,1,2$ can only be produced 
through three-gluon exchange given in Fig.1b, where the production is described with twist-3 GPD's obtained 
from $M_F^{\mu\nu\rho}$ by replacing $i f^{abc}$ in Eq.(\ref{GGPD}) by $d^{abc}$. 

\par  
We have derived the twist-2- and twist-3  amplitude of $h_c$.  At twist-2 $h_c$ is transversely polarized. The amplitude reads  
\begin{eqnarray} 
   {\mathcal T}_T &=& 16 g_s^2 A_P \epsilon^*_\nu (k) \epsilon^\rho (q) \epsilon_{\perp\mu\rho}   \int d x 
  \frac { \xi } { (\xi-x-i\varepsilon)^2 (\xi+x  -i\varepsilon) } \biggr [ g_\perp^{\mu\nu} {\mathcal F}_g 
     - i\epsilon_\perp^{\mu\nu} \tilde {\mathcal F}_g + 2 {\mathcal F}^{\mu\nu}_{Tg} \biggr ] ,  
\nonumber\\
     && A_P =e_Q  \sqrt{\frac{3}{4\pi N_c}} R'^* (0) (2 m_Q)^{-5/2},   
\label{T2hc}             
\end{eqnarray}
where $R' (0)$ is the derivative of the radial wave function at the origin. It is interesting to note that all twist-2 gluon GPD's are involved. But the production of $h_c$ can be difficult to be measured. 
At twist-3 $h_c$ is longitudinally polarized. The amplitude is:
\begin{eqnarray} 
{\mathcal T}_L &=& \frac{ 4 g_s^2 A_P}{m_Q} \int d x \biggr \{   \tilde \epsilon \cdot \Delta_\perp {\mathcal F}_g \biggr [ \frac{-2}{(\xi-x-i\varepsilon)(\xi+x -i\varepsilon)} +\xi \biggr ( 
\frac{ 5 x - 3\xi }{(\xi-x-i\varepsilon)(\xi+x-i\varepsilon)^3 } 
\nonumber\\
 && +\frac{\rho_L}{(\xi-x-i\varepsilon)^2 (\xi+x-i\varepsilon)} \biggr ) \biggr] 
+ i \xi  \epsilon\cdot \Delta_\perp \tilde{\mathcal  F} _g \biggr [\frac{2x -6 \xi }{(\xi-x-i\varepsilon)(\xi+x-i\varepsilon)^3 } 
\nonumber\\
&& 
 -\frac{\rho_L} {(\xi-x-i\varepsilon)^2 (\xi+x-i\varepsilon)} \biggr]
  - \xi F_T^{\mu\nu} \biggr [
    \frac{(10\xi-6 x )   \tilde \epsilon_\mu \Delta_{\perp\nu} }{(\xi-x-i\varepsilon)(\xi+x-i\varepsilon)^3 }  + \frac{\rho_L \ \epsilon_\mu \tilde \Delta_{\perp\nu} } {(\xi-x-i\varepsilon)^2 (\xi+x-i\varepsilon)} \biggr] \biggr \} 
\nonumber\\
  &&  +\frac{8 g_s^2 A_P \xi}{m_Q (1-\xi) } \int d x_1 d x_2 \biggr \{   2 \tilde \epsilon_\mu  G^\mu_\partial (x_1)\delta (x_1-x_2)  \frac{\xi-2 x_1 }{(\xi-x_1-i\varepsilon)^2 (\xi+ x_2 -i\varepsilon)^2} 
\nonumber\\
   && + \frac{  (x_1+x_2) \tilde \epsilon_\mu G_D^\mu (x_1,x_2)}{(\xi-x_2-i\varepsilon)(\xi+x_1-i\varepsilon) (\xi-x_1-i\varepsilon)(\xi+x_2 -i\varepsilon)} -\frac{4 \tilde \epsilon_\mu \tilde G_D^\mu (x_1,x_2)}{(\xi+x_2-i\varepsilon)^3}  
\nonumber\\
   && +4    \tilde \epsilon_\mu \tilde G_\partial^\mu (x_1) \delta (x_1-x_2) \frac{ x_1}{(\xi-x_1-i\varepsilon)^2 (\xi+x_1-i\varepsilon)^2}   -4 \tilde \epsilon_\mu \frac{\sum_q N_F^\mu (x_1,x_2) }{(2\xi+x_1-x_1-i\varepsilon)^3}  \biggr \},
\label{TLHC}             
\end{eqnarray}    
with $\tilde\epsilon^\mu =\epsilon_\perp^{\mu\nu} \epsilon_\nu$ and $\tilde\Delta^\mu =\epsilon_\perp^{\mu\nu} \Delta_\nu$.  Again, the amplitude here is suppressed only by the inverse of $m_Q$ in comparison with the twist-2 amplitude in Eq.(\ref{T2hc}). But this amplitude is in fact divergent.    

\par
Although GPD's can not be calculated perturbatively, their dependence on the renormalization scale 
$\mu$ can be calculated with perturbative theory if $\mu$ is large enough.    
The evolutions of twist-2 GPD's have been studied in \cite{RadEvo}. The asymptotic behaviors   
have been derived.  
In the limit of $\mu\to \infty$ the $x$-dependence of the GPD $E_g$ and $H_g$ contained in ${\mathcal F}_g$ can be found in \cite{RadEvo,BeRa}:  
\begin{equation} 
E_g (x,\xi) \sim \frac{4 C_F}{4 C_F + n_f}\frac{1}{\xi^5} (\xi-x)^2 (\xi+x)^2 \theta(\xi-\vert x\vert), \quad H_g (x,\xi) \sim \frac{4 C_F}{4 C_F + n_f}\frac{1}{\xi^5} (\xi-x)^2 (\xi+x)^2 \theta(\xi-\vert x\vert). 
\end{equation}
In the limit these two GPD's are zero for  $\vert x\vert > \xi$. Given such a dependence, the integrals like 
\begin{equation}
\int dx \frac{1}{(\xi + x -i\varepsilon)^3} \biggr \{ E_g(x,\xi),H_g(x,\xi)\biggr \}，
\label{IDIV}
\end{equation}
are divergent in the limit $ \mu \to \infty$. 
The last term in the first line 
in Eq.(\ref{TLHC}) contains such integrals  but at finite $\mu$.
One can calculate these integrals at $\mu\to \infty$ and obtain the results at finite $\mu$ 
from the evolution of the two GPD's. The evolutions can not eliminate such divergences. 
Therefore, the result in Eq.(\ref{TLHC}) are in fact divergent. The divergences 
comes form the momentum region where one of the two exchanged gluons 
in Fig.1a is soft. 
Such divergences are end-point ones, which 
are well-known in collinear factorization of exclusive processes with light-cone amplitudes of hadrons. 
For processes factorized with GPD's, the existence of such divergences involving GPD's has been first observed 
in $\rho$-meson production in\cite{MaPi}.  It is noted that in the twist-3 amplitude of $J/\psi$ or $\Upsilon$ in Eq.(\ref{T3}) there is no evidence of existing end-point singularities. One can expect that the factorization holds in this case.  

\par

\par 
To summarize: We have studied at twist-3 level the forward photoproduction of  a quarkonium, where the quarkonium can be  $J/\psi$ and  $h_c$.  We have classified the relevant twist-3 gluon GPD's and studied their properties.   
The produced quarkonium at twist-2 is transversely polarized. 
At twist-3 it is 
longitudinally polarized and the amplitude is power-suppressed.
Our result indicates that the twist-3 amplitude of $h_c$ contains end-point 
singularities which spoil the factorization.   
We find that the twist-3 amplitude 
is only suppressed by the inverse of the heavy quark mass. This has an important implication for extracting twist-2 gluon GPD's, if        
the polarization of the produced quarkonium is not observed. In this case, the contribution of the longitudinalxi- polarization can be significant 
and should be taken into account.  Experimentally, the studied processes can be already studied at JLab, and in future at EIC or After@LHC as
proposed in \cite{EIC} or \cite{AFTLHC}, respectively. 
 
\par\vskip20pt
\noindent
{\bf Acknowledgments}
\par
The work is supported by National Nature
Science Foundation of P.R. China(No.11675241 and  No.11747601). The partial support from the CAS center for excellence in particle 
physics(CCEPP) is acknowledged.

\par

\end{document}